\documentclass[12pt]{article}
\usepackage{amsmath,url}
\usepackage{graphicx}
\input{epsf} 
\providecommand{\href}[2]{#2}

\def\gtap{\raisebox{-.6ex}{\rlap{$\,\sim\,$}} \raisebox{.4ex}{$\,>\,$}}

\newcommand\as{\alpha_{\mathrm{S}}} 
\newcommand\f[2]{\frac{#1}{#2}} 
 
\def\to{\rightarrow}

\begin{document} 
\begin{titlepage}
\renewcommand{\thefootnote}{\fnsymbol{footnote}}
\begin{flushright}
ZU-TH 21/13
\end{flushright}
\vspace*{2cm}

\begin{center}
{\Large \bf $Z\gamma$ production at hadron colliders in NNLO QCD}
\end{center}

\par \vspace{2mm}
\begin{center}
{\bf Massimiliano Grazzini\footnote{On leave of absence from INFN, Sezione di Firenze, Sesto Fiorentino, Florence, Italy.}, Stefan Kallweit, Dirk Rathlev}
and {\bf Alessandro Torre}
\vspace{5mm}

Institut f\"ur Theoretische Physik, Universit\"at Z\"urich, CH-8057 Z\"urich, Switzerland

\vspace{5mm}

\end{center}

\par \vspace{2mm}
\begin{center} {\large \bf Abstract} \end{center}
\begin{quote}
\pretolerance 10000

We consider the production of $Z\gamma$ pairs at hadron colliders.
We report on the first complete and fully differential
computation of radiative corrections at next-to-next-to-leading order
in QCD perturbation theory. We present selected numerical results for $pp$ collisions at 7 TeV and compare them
to available LHC data. We find that the impact of the
NNLO QCD corrections on the fiducial cross section ranges between
$4$ and $15\%$, depending on the applied cuts. 

\end{quote}

\vspace*{\fill}
\begin{flushleft}
September 2013

\end{flushleft}
\end{titlepage}

\setcounter{footnote}{1}
\renewcommand{\thefootnote}{\fnsymbol{footnote}}

The production of vector-boson pairs is a crucial process
for physics studies within and beyond the Standard Model (SM).
In particular, the production of neutral vector-boson pairs, like
$Z\gamma$, is well suited to search for anomalous couplings.
Despite the non-Abelian structure
of the $SU(2)_L\otimes U(1)_Y$ gauge group, which entails
self-interactions of the gauge bosons,
a $ZZ\gamma$ coupling in the SM is not allowed
as the $Z$ boson is electrically neutral.
The production of $Z\gamma$
at hadron colliders is thus dominated by diagrams
in which the photon is radiated
either off an initial-state quark or the final-state leptons.
A non-zero $ZZ\gamma$ coupling would be a clear signal for new physics.
 
The production of $Z\gamma$ is also
a background for Higgs boson searches.
The Higgs decay into $Z\gamma$ final states in the SM is a rare loop-induced
process with a very small branching ratio.
However, this is not necessarily the case in extensions of the SM, so
the $Z\gamma$ rates can be used to discriminate
between new-physics models.

Recent measurements of the $Z\gamma$ cross section carried out at the Tevatron Run II and at the LHC have been reported in Refs.~\cite{Aaltonen:2011zc,Aad:2013izg,Chatrchyan:2013fya}.

When considering the $Z\gamma$ final state,
besides the {\it direct} production in the hard subprocess,
the photon can also be
produced through the {\it fragmentation} of a QCD parton, and the evaluation of the ensuing contribution to the cross section
requires the knowledge of a non-perturbative photon fragmentation function,
which typically has large uncertainties.
The fragmentation contribution is significantly suppressed by the photon isolation criteria
that are necessarily
applied in hadron-collider experiments in order to suppress the large backgrounds.
The {\it standard cone} isolation, which is usually applied
in the experiments, suppresses a large fraction of the fragmentation component.
The {\it smooth cone} isolation completely suppresses the fragmentation contribution \cite{Frixione:1998jh}, but it is difficult to be implemented experimentally.

The status of theoretical predictions for $Z\gamma$ production at
hadron colliders is as follows.
The $Z\gamma$ cross section is known in NLO QCD \cite{Ohnemus:1992jn},
including the leptonic decay of the $Z$ boson \cite{DeFlorian:2000sg}.
The loop-induced gluon fusion contribution, which is formally
next-to-next-to-leading order (NNLO), has been computed in Ref.~\cite{Ametller:1985di},
and the leptonic decay of the $Z$ boson,
together with the gluon-induced tree level NNLO contributions, have been added in Ref.~\cite{Adamson:2002rm}.
The NLO calculation, including photon radiation from the final-state leptons, the loop-induced
gluon contribution and the photon fragmentation at LO
have been implemented into the general purpose numerical program {\sc MCFM} \cite{Campbell:2011bn}.
Electroweak (EW) corrections to $Z\gamma$ production
have been computed in Ref.~\cite{Hollik:2004tm}.

In this Letter we report on the first complete computation of $pp\to Z\gamma+X$ in NNLO QCD.
We note that the notation ``$Z\gamma$'' is misleading, as it suggests
the production of an on-shell $Z$ boson plus a photon,
followed by a factorized decay of the $Z$ boson.
Instead, we actually compute the NNLO corrections to the process
$pp\to l^+l^-\gamma+X$, where the lepton pair $l^+l^-$ is
produced either by a $Z$ boson or a virtual photon,
and we consistently
include the contributions in which the final-state photon is radiated from the leptons.
The NNLO computation requires the evaluation of the tree-level
scattering amplitudes with two additional (unresolved) partons, of the one-loop amplitudes with one additional parton \cite{Bern:1997sc,Campbell:2012ft}, and of the one-loop squared and
two-loop corrections to the Born subprocess $q{\bar q}\to l^+l^-\gamma$.
In our computation the required tree-level and one-loop
amplitudes are obtained by using
the {\sc OpenLoops} generator \cite{Cascioli:2011va},
which is based on a new numerical approach for
the recursive construction of cut-opened loop diagrams.
The {\sc OpenLoops} generator employs
the Denner--Dittmaier algorithm for the numerically stable evaluation of
tensor integrals~\cite{Denner:2005nn} and allows a fast
evaluation of tree-level and one-loop amplitudes within the SM.

The two-loop correction to the Born process in which the photon
is radiated off the final state leptons is available since long time \cite{Matsuura:1988sm}.
The last missing contribution,
the genuine two-loop correction to the $Z\gamma$ amplitude,
has recently been presented in Ref.~\cite{Gehrmann:2011ab}.

The implementation of the various scattering amplitudes in a complete NNLO calculation is
a highly non-trivial task due to the presence of infrared (IR) singularities at
intermediate stages of the calculation that prevent a straightforward implementation of numerical techniques.
The $q_T$ subtraction formalism \cite{Catani:2007vq} is
a method to handle and cancel these singularities at the NNLO.
The formalism applies to the
production of a colourless high-mass system $F$
in generic hadron collisions and has been applied to the computation of
NNLO corrections to several hadronic processes \cite{Catani:2007vq,Catani:2009sm}.
According to the $q_T$ subtraction method \cite{Catani:2007vq}, the $pp\to F+X$ cross section can be written as
\begin{equation}
\label{main}
d{\sigma}^{F}_{(N)NLO}={\cal H}^{F}_{(N)NLO}\otimes d{\sigma}^{F}_{LO}
+\left[ d{\sigma}^{F+{\rm jets}}_{(N)LO}-
d{\sigma}^{CT}_{(N)LO}\right]\;\; ,
\end{equation}
where $d{\sigma}^{F+{\rm jets}}_{(N)LO}$ represents the cross section for the
production of the system $F$ plus jets at (N)LO accuracy, and can be evaluated with
any available version of the NLO subtraction formalism.
The (IR subtraction) counterterm $d{\sigma}^{CT}_{(N)LO}$
is obtained from the resummation program of the logarithmically-enhanced
contributions to $q_T$ distributions \cite{Bozzi:2005wk}.  
The `coefficient' ${\cal H}^{F}_{(N)NLO}$, which also compensates for the subtraction
of $d{\sigma}^{CT}_{(N)LO}$,
corresponds to the (N)NLO truncation of the process-dependent perturbative function
\begin{equation}
{\cal H}^{F}=1+\f{\as}{\pi}\,
{\cal H}^{F(1)}+\left(\f{\as}{\pi}\right)^2
{\cal H}^{F(2)}+ \dots \;\;.
\end{equation}
The NLO calculation  of $d{\sigma}^{F}$ 
requires the knowledge
of ${\cal H}^{F(1)}$, and the NNLO calculation also requires ${\cal H}^{F(2)}$.

The general structure of ${\cal H}^{F(1)}$
is known \cite{deFlorian:2001zd}: 
${\cal H}^{F(1)}$ is obtained from the process-dependent scattering
amplitudes by using a process-independent relation.
Exploiting the explicit results of ${\cal H}^{F(2)}$ for Higgs
\cite{Catani:2011kr} and vector boson \cite{Catani:2012qa} 
production,
the process-independent relation of 
Ref.~\cite{deFlorian:2001zd} has been extended to the calculation of the NNLO coefficient 
${\cal H}^{F(2)}$ \cite{Catani:2013tia}.
We have performed our fully-differential NNLO calculation of $Z\gamma$ production
according to Eq.~(\ref{main}), starting from a computation
of the $d{\sigma}^{Z\gamma+{\rm jets}}_{NLO}$ cross section with the dipole subtraction method
\cite{Catani:1996jh}\footnote{An independent calculation of 
$d{\sigma}^{Z\gamma+{\rm jets}}_{NLO}$ was performed in 
Ref.~\cite{Campbell:2012ft}.}.

The NNLO computation is encoded
in a parton-level
Monte Carlo program that allows us
to apply arbitrary IR safe cuts on the $l^+l^-\gamma$ final state
and the associated jet activity. 
The program is based on the fully automatized framework developed in the calculations of Ref.~\cite{Denner:2010jp};
it generates each involved phase-space in a multi-channel approach and constructs
the required Catani--Seymour dipoles including extra phase-space mappings according to their modified kinematics. Additionally, importance-sampling techniques are applied to further improve
the convergence in phase-space regions where $q_T\gtap 0$.

The present formulation of the $q_T$ subtraction formalism \cite{Catani:2007vq}
is limited to the production of colourless systems $F$ and, hence, it does not
allow us to deal with the 
parton fragmentation subprocesses.
Therefore, we consider only direct photons, and 
we rely on the smooth cone isolation criterion \cite{Frixione:1998jh}.
Considering a cone of radius $r=\sqrt{(\Delta \eta)^2+(\Delta \phi)^2}$ around
the photon, we require that the total amount of hadronic (partonic) transverse energy $E_T$ 
inside the cone is smaller than $E_{T}^{\rm max}(r)$,
\begin{equation}
E_{T}^{\rm max}(r) \equiv  \epsilon_\gamma \,p_T^\gamma \left(\frac{1-\cos r}{1- \cos R}\right)^n \, ,
\end{equation}
where $p_T^\gamma$ is the photon transverse momentum; the isolation criterion
$E_T < E_{T}^{\rm max}(r)$ has to be fulfilled for all cones with $r\leq R$.
Unless stated otherwise,
the results presented in this Letter are obtained
with $\epsilon_\gamma=0.5$, $n=1$ and $R=0.4$.

In the following we present a selection of our numerical results for $pp$ collisions with $\sqrt{s}=7$ TeV.
As for the electroweak couplings, we use the so called $G_\mu$ scheme,
where the input parameters are $G_F$, $m_W$, $m_Z$. In particular we 
use the values
$G_F = 1.16639\times 10^{-5}$~GeV$^{-2}$, $m_W=80.398$ GeV,
$m_Z = 91.1876$~GeV, $\Gamma_Z=2.4952$~GeV.
We use the MSTW 2008 \cite{Martin:2009iq} sets of parton distributions, with
densities and $\as$ evaluated at each corresponding order
(i.e., we use $(n+1)$-loop $\as$ at N$^n$LO, with $n=0,1,2$),
and we consider $N_f=5$ massless quarks/antiquarks and gluons in 
the initial state. The default
renormalization ($\mu_R$) and factorization ($\mu_F$) scales are set to
$\mu_R=\mu_F=\mu_0\equiv\sqrt{m_Z^2+(p_T^{\gamma})^2}$.

We first consider the selection cuts that are applied by the ATLAS collaboration \cite{Aad:2013izg}.
We require the photon to have a transverse momentum $p_T^\gamma>15$ GeV and pseudorapidity $|\eta^\gamma|<2.37$. The charged leptons are required to have $p_T^l>25$ GeV and $|\eta^l|<2.47$, and their invariant mass $m_{ll}$ must fulfil $m_{ll}>40$ GeV.
We require the separation in rapidity and azimuth $\Delta R$ between
the leptons and the photon to be $\Delta R(l,\gamma)>0.7$.
Jets are reconstructed with the anti-$k_T$ algorithm \cite{Cacciari:2008gp} with radius parameter $D=0.4$. A jet must have $E_T^{\rm jet}>30$ GeV and $|\eta^{\rm jet}|<4.4$. We require the separation $\Delta R$ between the
leptons (photon) and the jets to be $\Delta R(l/\gamma,{\rm jet})>0.3$.
Our results for the corresponding cross sections\footnote{Throughout the paper,
the errors on the values of the cross sections and the error bars in the histograms refer to an estimate of the numerical uncertainties in our calculation.} are $\sigma_{LO}=850.7\pm 0.2$ fb, $\sigma_{NLO}=1226.2\pm 0.4$ fb and $\sigma_{NNLO}=1305\pm 3$ fb.
The NNLO corrections increase the NLO result by $6\%$. The loop-induced $gg$ contribution
amounts to $8\%$ of the ${\cal O}(\as^2)$
correction and thus to less than $1\%$ of $\sigma_{NNLO}$.

We have studied the dependence of our results on the renormalization and factorization scales. We find that, when the scales
are varied around the default scale $\mu_0$ in the same direction (i.e. setting $\mu_R=\mu_F=a\mu_0$ and
varying $a$ between 0.5 and 2), the effect at NLO and NNLO
is essentially negligible.
By following Ref.~\cite{Campbell:2011bn} and setting $\mu_R=a\mu_0$ and $\mu_F=\mu_0/a$,
the effect is $-5\%$ ($+4\%$) at NLO and $-2\%$ ($+2\%$) at NNLO for $a=0.5$ ($a=2$), respectively.

Our results can be compared with the ATLAS data \cite{Aad:2013izg}\footnote{The comparison with the experimental data should be taken with a grain of salt.
The photon isolation used in the experiment is different from ours. Furthermore, our predictions should be corrected for the differences between the parton-level and hadron-level definitions of jets and photons.}.
The fiducial cross section measured by ATLAS is $\sigma=1.31\pm 0.02~{\rm (stat)}\pm 0.11~{\rm (syst)} \pm 0.05~{\rm (lumi)}$ pb.
The NNLO effects improve the agreement of the QCD prediction with the data, which, however, still have relatively large uncertainties.

\begin{figure}[htb]
\begin{center}
\begin{tabular}{c}
\epsfxsize=10truecm
\epsffile{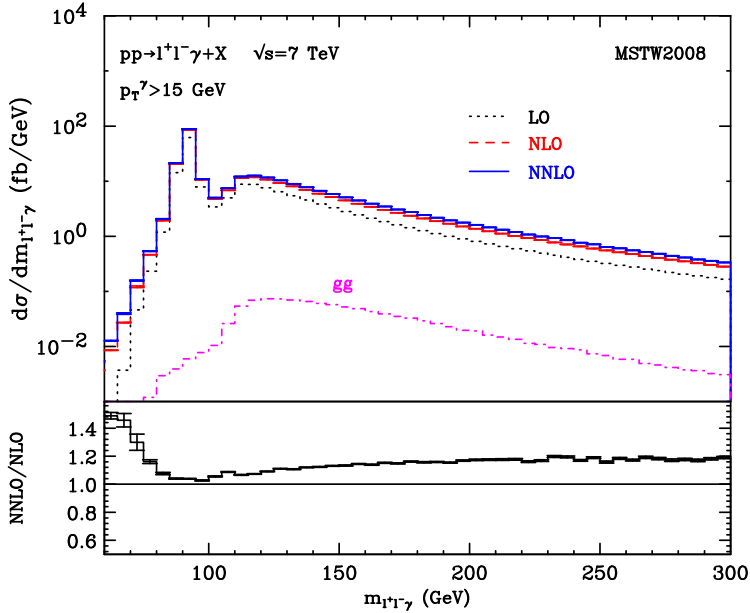}\\
\end{tabular}
\end{center}
\caption{\label{fig:mZg}
{\em Invariant mass distribution of the $l^+l^-\gamma$ system at LO (dots), NLO (dashes), NNLO (solid). The loop-induced $gg$ contribution is also shown for comparison. The lower panel shows the ratio NNLO/NLO.}}
\end{figure}

In Fig.~\ref{fig:mZg} we study the impact of QCD radiative corrections on the invariant-mass distribution of the $l^+l^-\gamma$ system. The lower panel shows the ratio NNLO/NLO.
We see that the impact of NNLO corrections is not uniform over the range of $m_{l^+l^-\gamma}$.
NNLO corrections are relatively small in the region where the cross section is higher,
and larger above $150$ GeV. The LO distribution has a kinematical boundary at $m_{l^+l^-\gamma}\sim 66$ GeV, and the region below this boundary receives contributions only beyond LO.
We also note that the invariant-mass region below the $Z$ peak is the one in which NNLO corrections are more significant, but it marginally contributes to the cross section.

\begin{figure}[htb]
\begin{center}
\begin{tabular}{c}
\epsfxsize=10truecm
\epsffile{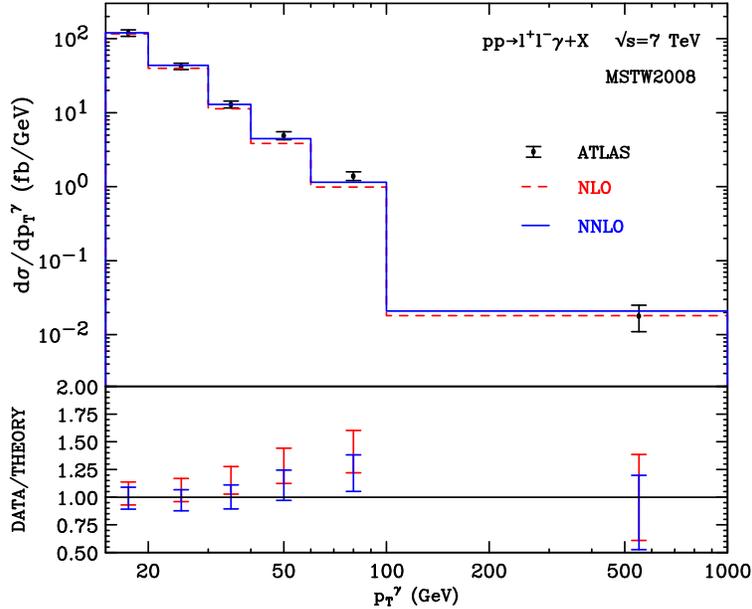}\\
\end{tabular}
\end{center}
\caption{\label{fig:Etgamma}
{\em Transverse momentum spectrum of the photon at NLO and NNLO compared with ATLAS data. The lower panel shows the ratio DATA/THEORY.}}
\end{figure}

In Fig.~\ref{fig:Etgamma} we consider the $p_T$ distribution of the photon, and we present a comparison of the NLO and NNLO theoretical predictions with the ATLAS data (the bin sizes are chosen so as to match those adopted in Ref.~\cite{Aad:2013izg}). We see that the data agree with the NLO and NNLO theoretical predictions within the uncertainties, and that the NNLO corrections slightly improve this agreement.
We should not forget, however, that EW corrections affect the tail of the $p_T^\gamma$ distribution in a significant way and act in the opposite direction \cite{Hollik:2004tm}.

\begin{figure}[htb]
\begin{center}
\begin{tabular}{c}
\epsfxsize=10truecm
\epsffile{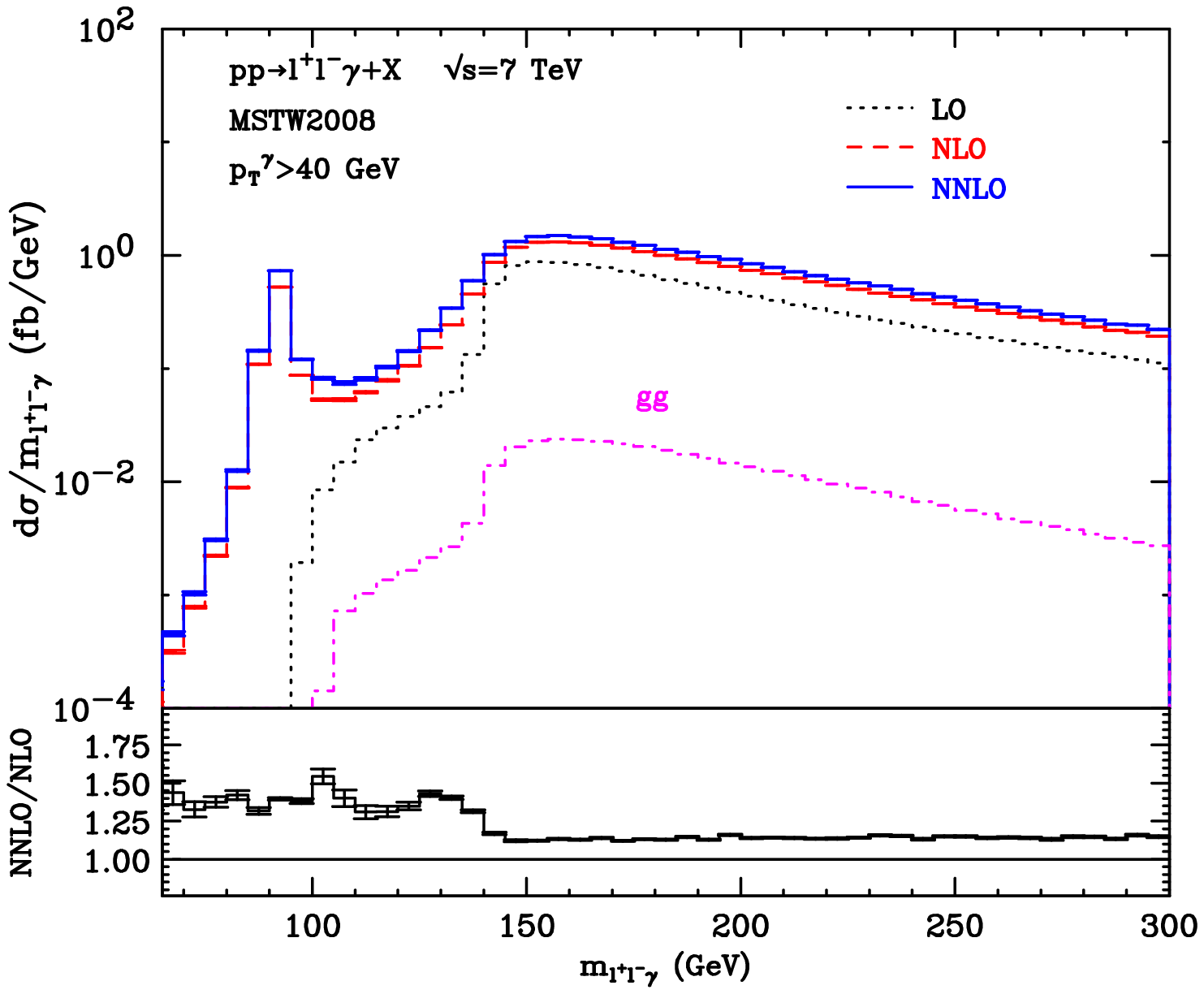}\\
\end{tabular}
\end{center}
\caption{\label{fig:mZg40}
{\em As in Fig.~\ref{fig:mZg} but for $p_T^\gamma>40$ GeV.}
}
\end{figure}

ATLAS also considers an additional set up with $p_T^\gamma>40$ GeV,
for which, however, the measured fiducial cross section is not provided.
In this case our corresponding cross sections are $\sigma_{LO}=77.48\pm 0.06$ fb, $\sigma_{NLO}=132.89\pm 0.07$ fb and $\sigma_{NNLO}=152.5\pm 0.5$ fb. The impact of the NNLO corrections is about $15\%$ with respect to NLO. The increased impact of NNLO corrections compared to the $p_T^\gamma>15$ GeV case
can be understood by studying the invariant mass distribution in Fig.~\ref{fig:mZg40}.
With $p_T^\gamma>40$ GeV
the LO boundary moves to $m_{l^+l^-\gamma}\sim 97$ GeV, and
the phase-space region below the boundary, which opens up
beyond LO, includes the
$Z$ peak, and significantly contributes to the cross section.
Moreover the region immediately above the $Z$ peak shows relatively large NLO and NNLO corrections.

We have also considered the selection cuts applied by the CMS collaboration \cite{Chatrchyan:2013fya}. They require the photon to have $p_T^\gamma>15$ GeV and pseudorapidity $|\eta^\gamma|<2.5$. The charged leptons are required to have $p_T^l>20$ GeV, $|\eta^l|<2.5$, and $m_{ll}>50$ GeV. The lepton--photon separation is $\Delta R(l,\gamma)>0.7$.
The photon-isolation parameters that we use in this case
are $\epsilon_\gamma=0.05$ and $R=0.3$.
Our corresponding results are $\sigma_{LO}=1333.6\pm 0.2$ fb, $\sigma_{NLO}=1843.8 \pm 0.7$ fb and $\sigma_{NNLO}=1917 \pm 8$ fb.
The impact of NNLO corrections on the NLO result is about $4\%$.
A direct comparison to CMS data is not possible because CMS does not provide the measured fiducial cross section.

We note that the photon isolation parameters used by CMS are rather different from those used by ATLAS.
To estimate the impact of the different isolation parameters on the results,
we have repeated our calculation for the CMS selection cuts by using
the isolation parameters of the ATLAS analysis, i.e. $\epsilon_\gamma=0.5$ and $R=0.4$.
We find that the NLO and NNLO cross sections are rather stable, since they
increase only by $0.2\%$ and $1\%$, respectively.

We have illustrated 
the first calculation of the cross section for $Z\gamma$ production at the LHC
up to NNLO in QCD perturbation theory. Our computation is implemented in a numerical program that allows us to apply arbitrary kinematical cuts on the final state leptons and photon and on the associated jet activity. For the selection cuts typically applied by the ATLAS and CMS collaborations, we find that the impact of NNLO corrections is moderate, and ranges between $4$ and $15\%$. The impact of NNLO corrections may be larger in some
kinematical regions.

\noindent {\bf Acknowledgements.}
It is our pleasure to thank Fabio Cascioli, Philipp Maierh{\"o}fer and Stefano Pozzorini for their continuous support with the {\sc OpenLoops} generator, and
Lorenzo Tancredi for helpful discussions on the results of Ref.~\cite{Gehrmann:2011ab}.
We also thank Stefano Catani and Daniel de Florian for comments on the manuscript.
We are grateful to Nicolas Chanon and Mauro Donega for useful discussions on photon isolation in CMS.
This research was supported in part by the Swiss National Science Foundation (SNF) under contracts CRSII2-141847, 200021-144352 and by 
the Research Executive Agency (REA) of the European Union under the Grant Agreement number PITN-GA-2010-264564 ({\it LHCPhenoNet}).


\begin{thebibliography}{99}

\bibitem{Aaltonen:2011zc}
  T.~Aaltonen {\it et al.}  [CDF Collaboration],
  Phys.\ Rev.\ Lett.\  {\bf 107} (2011) 051802;
  V.~M.~Abazov {\it et al.}  [D0 Collaboration],
  Phys.\ Rev.\ D {\bf 85} (2012) 052001.

\bibitem{Aad:2013izg}
  G.~Aad {\it et al.}  [ATLAS Collaboration],
  Phys.\ Rev.\ D {\bf 87} (2013) 112003.

\bibitem{Chatrchyan:2013fya}
  S.~Chatrchyan {\it et al.}  [CMS Collaboration],
report CMS-EWK-11-009, CERN-PH-EP-2013-108,  arXiv:1308.6832 [hep-ex].

\bibitem{Frixione:1998jh}
  S.~Frixione,
  Phys.\ Lett.\ B {\bf 429} (1998) 369.

\bibitem{Ohnemus:1992jn}
  J.~Ohnemus,
  Phys.\ Rev.\ D {\bf 47} (1993) 940;
  U.~Baur, T.~Han and J.~Ohnemus,
  Phys.\ Rev.\ D {\bf 57} (1998) 2823.

\bibitem{DeFlorian:2000sg}
  D.~de Florian and A.~Signer,
  Eur.\ Phys.\ J.\ C {\bf 16} (2000) 105.

\bibitem{Ametller:1985di}
  L.~Ametller, E.~Gava, N.~Paver and D.~Treleani,
  Phys.\ Rev.\ D {\bf 32} (1985) 1699;
  J.~J.~van der Bij and E.~W.~N.~Glover,
  Phys.\ Lett.\ B {\bf 206} (1988) 701.

\bibitem{Adamson:2002rm}
  K.~L.~Adamson, D.~de Florian and A.~Signer,
  Phys.\ Rev.\ D {\bf 67} (2003) 034016.

\bibitem{Campbell:2011bn}
  J.~M.~Campbell, R.~K.~Ellis and C.~Williams,
  JHEP {\bf 1107} (2011) 018.


\bibitem{Hollik:2004tm}
  W.~Hollik and C.~Meier,
  Phys.\ Lett.\ B {\bf 590} (2004) 69;
  E.~Accomando, A.~Denner and C.~Meier,
  Eur.\ Phys.\ J.\ C {\bf 47} (2006) 125.

\bibitem{Bern:1997sc}
  Z.~Bern, L.~J.~Dixon and D.~A.~Kosower,
  Nucl.\ Phys.\ B {\bf 513} (1998) 3.

\bibitem{Campbell:2012ft}
  J.~M.~Campbell, H.~B.~Hartanto and C.~Williams,
  JHEP {\bf 1211} (2012) 162.


\bibitem{Cascioli:2011va}
  F.~Cascioli, P.~Maierh{\"o}fer and S.~Pozzorini,
  Phys.\ Rev.\ Lett.\  {\bf 108} (2012) 111601.

\bibitem{Denner:2005nn}
  A.~Denner and S.~Dittmaier,
  Nucl.\ Phys.\ B {\bf 734} (2006) 62,
  Nucl.\ Phys.\ B {\bf 844} (2011) 199.

\bibitem{Matsuura:1988sm}
  T.~Matsuura, S.~C.~van der Marck and W.~L.~van Neerven,
  Nucl.\ Phys.\ B {\bf 319} (1989) 570.

\bibitem{Gehrmann:2011ab}
  T.~Gehrmann and L.~Tancredi,
  JHEP {\bf 1202} (2012) 004.

\bibitem{Catani:2007vq}
  S.~Catani and M.~Grazzini,
  Phys.\ Rev.\ Lett.\  {\bf 98} (2007) 222002.


\bibitem{Catani:2009sm}
  S.~Catani, L.~Cieri, G.~Ferrera, D.~de Florian and M.~Grazzini,
  Phys.\ Rev.\ Lett.\  {\bf 103} (2009) 082001;
  G.~Ferrera, M.~Grazzini and F.~Tramontano,
  Phys.\ Rev.\ Lett.\  {\bf 107} (2011) 152003;
  S.~Catani, L.~Cieri, D.~de Florian, G.~Ferrera and M.~Grazzini,
  Phys.\ Rev.\ Lett.\  {\bf 108} (2012) 072001.


\bibitem{Bozzi:2005wk}
  G.~Bozzi, S.~Catani, D.~de Florian and M.~Grazzini,
  Nucl.\ Phys.\ B {\bf 737} (2006) 73.



\bibitem{deFlorian:2001zd}
  D.~de Florian and M.~Grazzini,
  Nucl.\ Phys.\ B {\bf 616} (2001) 247.


\bibitem{Catani:2011kr}
  S.~Catani and M.~Grazzini,
  Eur.\ Phys.\ J.\ C {\bf 72} (2012) 2013
   [Erratum-ibid.\ C {\bf 72} (2012) 2132].

\bibitem{Catani:2012qa}
  S.~Catani, L.~Cieri, D.~de Florian, G.~Ferrera and M.~Grazzini,
  Eur.\ Phys.\ J.\ C {\bf 72} (2012) 2195.

\bibitem{Catani:2013tia}
  S.~Catani, L.~Cieri, D.~de Florian, G.~Ferrera and M.~Grazzini, report ZU-TH 25/13,
  arXiv:1311.1654 [hep-ph].



\bibitem{Catani:1996jh}
  S.~Catani and M.~H.~Seymour,
  Phys.\ Lett.\ B {\bf 378} (1996) 287,
  Nucl.\ Phys.\ B {\bf 485} (1997) 291
   [Erratum-ibid.\ B {\bf 510} (1998) 503].

\bibitem{Denner:2010jp}
  A.~Denner, S.~Dittmaier, S.~Kallweit and S.~Pozzorini,
  Phys.\ Rev.\ Lett.\  {\bf 106} (2011) 052001,
  JHEP {\bf 1210} (2012) 110;
  A.~Denner, L.~Hosekova and S.~Kallweit,
  Phys.\ Rev.\ D {\bf 86} (2012) 114014.


\bibitem{Martin:2009iq}
  A.~D.~Martin, W.~J.~Stirling, R.~S.~Thorne and G.~Watt,
  Eur.\ Phys.\ J.\ C {\bf 63} (2009) 189.

\bibitem{Cacciari:2008gp}
  M.~Cacciari, G.~P.~Salam and G.~Soyez,
  JHEP {\bf 0804} (2008) 063.


\end{thebibliography}
\end{document}